\documentclass[3p]{elsarticle}
   
\biboptions{sort&compress}

 \pdfoutput=1 
 
 \makeatletter
 \def\ps@pprintTitle{%
  \let\@oddhead\@empty
  \let\@evenhead\@empty
 \def\@oddfoot{}%
  \let\@evenfoot\@oddfoot}
 \makeatother

\usepackage{color}
\usepackage{array}
\usepackage{graphicx}

\usepackage[english]{babel}
\usepackage[utf8]{inputenc}
\usepackage[tbtags]{amsmath}
\usepackage{amsfonts}
\usepackage{amssymb}
\usepackage{slashed}
\usepackage{graphicx}
\usepackage[T1]{fontenc}
\usepackage{ae}
\usepackage{color}
\usepackage[colorlinks=true,linkcolor=darkred,citecolor=darkred,urlcolor=darkred, pdfborder={0 0 0}]{hyperref}

\newcommand{\eq}[1]{Eq.~(\ref{#1})}

\newcommand{\Eq}[1]{Eq.~(\ref{#1})}

\definecolor{gray}{gray}{0.1}
\definecolor{darkred}{rgb}{0.6471, 0.1098, 0.1882} 
\definecolor{darkblue}{rgb}{0.1098, 0.1882, 0.6471}

\def\beqn#1{\begin{equation}\label{#1}}
\def\eeqn{\end{equation}}

\def\beqa#1{\begin{eqnarray}\label{#1}}
\def\eeqa{\end{eqnarray}}

\def\gsim{\raise0.3ex\hbox{$\;>$\kern-0.75em\raise-1.1ex\hbox{$\sim\;$}}}
\def\lsim{\raise0.3ex\hbox{$\;<$\kern-0.75em\raise-1.1ex\hbox{$\sim\;$}}}

\newcommand{\gev}{\,\mathrm{GeV}}
\newcommand{\tev}{\,\mathrm{TeV}}
\newcommand{\mev}{\,\mathrm{MeV}}
\newcommand{\kev}{\,\mathrm{keV}}

\newcommand{\nn}{\nonumber}

\def\beqn#1{\begin{equation}\label{#1}}
\def\eeqn{\end{equation}}

\def\beqa#1{\begin{eqnarray}\label{#1}}
\def\eeqa{\end{eqnarray}}

\newcommand{\irrep}[1]{\ensuremath{{\bf{#1}}}}

\def\vev#1{\left\langle #1\right\rangle}

\def\z2{\mathbb{Z}_2}
\def\zn{\mathbb{Z}_N}

\newcommand{\eqn}[1]{eq.~(\ref{#1})}
\newcommand{\eqns}[2]{eqs.~(\ref{#1})-(\ref{#2})}

\newcommand{\Eqn}[1]{Eq.~(\ref{#1})}

\newcommand{\ten}{ \irrep{10}}

\newcommand{\sixtn}{ \irrep{16}}
\newcommand{\bsixtn}{ \irrep{\overline{16}}}
\newcommand{\hts}{ \irrep{126}}
\newcommand{\bhts}{ \irrep{\overline{126}}}
\newcommand{\fv}{ \irrep{45}}

\newcommand{\sufv}{ \irrep{5}}
\newcommand{\bsufv}{ \irrep{\bar 5}}

\newcommand{\lL}{\mathcal{\scriptscriptstyle L}}
\newcommand{\lR}{\mathcal{\scriptscriptstyle R}}
\newcommand{\lvec}[1]{\reflectbox{\ensuremath{\vec{\reflectbox{\ensuremath{#1}}}}}}

\begin{document}

 \baselineskip 16pt

\title{Dark Matter from the vector of SO(10)}

 \author[]{Sofiane M. Boucenna}
 \ead{boucenna@lnf.infn.it}
 \author[]{Martin B. Krauss}
 \ead{martin.krauss@lnf.infn.it}
 \author[]{Enrico Nardi}
 \ead{enrico.nardi@lnf.infn.it}
 \address{ INFN, Laboratori Nazionali di Frascati, C.P. 13, 100044 Frascati, Italy}
\date{\today}


\begin{abstract}
  $SO(10)$ grand unified theories can ensure the stability of new particles
   in terms of the gauge group structure itself, and in this
  respect are well suited to accommodate dark matter (DM) candidates
  in the form of new stable massive particles.  We introduce new
  fermions in two vector $\ten$ representations.  When $SO(10)$ is
  broken to the standard model by a minimal $\fv +\bhts+\ten$ scalar
  sector with
  $SU(3)_C \otimes SU(2)_L \otimes SU(2)_R\otimes U(1)_{B-L} $ as
  intermediate symmetry group , the resulting lightest new states are
  two Dirac fermions corresponding to combinations of the neutral
  members of the $SU(2)_L$ doublets in the $\ten$'s, which get
  splitted in mass by loop corrections involving $W_R$. The resulting
  lighter mass eigenstate is stable, and has only non-diagonal
  $Z_{L,R}$ neutral current couplings to the heavier neutral state.
  Direct detection searches are evaded if the mass splitting is
  sufficiently large to suppress kinematically inelastic
  light-to-heavy scatterings.  By requiring that this condition is
  satisfied, we obtain the upper limit $M_{W_R}\lsim 25$ TeV.
  \end{abstract}

\maketitle

\section{Introduction} 
A plethora of astrophysical and cosmological observations have firmly
established that non-baryonic dark matter (DM) must exist in our
Universe, and contribute to the overall cosmological energy density
about five times more than ordinary matter.  However, none of the
particles of the standard model (SM) can account for the DM, which
therefore constitutes a clear hint of new physics.  Colorless,
electrically neutral and weakly interacting massive particles 
with mass in the GeV-TeV range are ubiquitous in new physics models,
and appear to be well suited to reproduce quantitatively the measured
DM energy density if their stability on cosmological time scales
can be  ensured. \\
From the model-building point of view, DM stability is most commonly
enforced by assuming some suitable symmetry that forbids its decay
into lighter SM particles. For example, in supersymmetric models this
role is played by R-parity that stabilizes the lightest supersymmetric
state, in universal extra dimensional models conservation of
Kaluza-Klein parity ensures that the lightest Kaluza-Klein state
remains stable~\cite{Servant:2002aq}, $T$-parity stabilizes the
lightest $T$-odd particle in the littlest Higgs
model~\cite{Cheng:2004yc}, suitable $\z2$ parities play the same role
e.g. in the scotogenic model \cite{Ma:2006km,Farzan:2012sa}, in the
inert doublet
model~\cite{Deshpande:1977rw,Barbieri:2006dq,LopezHonorez:2006gr}, and
in several other cases.  Often these stabilizing symmetries are just
imposed by hand on the low energy Lagrangian, and it is certainly 
more satisfactory when their origin can be traced back
to some high energy completion of the model in question. A
plausible way to generate unbroken discrete $\zn$ symmetries relies
on assuming extra gauged $U(1)$ Abelian factors which are only broken
by order parameters carrying $N$ units of the $U(1)$
charge~\cite{Krauss:1988zc} (see also~\cite{Hamaguchi:1998wm,%
Hamaguchi:1998nj,Martin:1992mq,Batell:2010bp}).
Such a mechanism renders grand unified theories (GUTs) based on gauge
groups of rank larger than four particularly interesting, since they
contain extra Cartan generators besides the $2+1+1$ of the
$SU(3)_C\times SU(2)_L\times U(1)_Y$ SM gauge group that, when broken
by vacuum expectation values (vevs) of scalars in appropriate
representations, yield discrete $\zn$ symmetries which
inherit all the good properties of the parent local gauge symmetry.
In particular, this type of symmetries remain protected from gravity
induced symmetry breaking
effects~\cite{Banks:1989zw,Giddings:1988cx,Coleman:1988tj,%
  Gilbert:1989nq} which, although suppressed by the Planck scale, could 
jeopardize DM stability~\cite{Mambrini:2015sia,Boucenna:2012rc}.
 
One of the most interesting GUT groups that allows to preserve at low
energies an unbroken discrete gauge parity is the rank five group
$SO(10)$.  As is well known, $SO(10)$ has many theoretically appealing
properties: it unifies all SM fermions in a single $\mathbf{16}$
dimensional irreducible representation including one right handed (RH)
neutrino, it can explain the suppression of neutrino masses via the
seesaw
mechanism~\cite{Minkowski:1977sc,Yanagida:1979as,Glashow:1979nm,%
  GellMann:1980vs,Mohapatra:1979ia}, it allows for gauge coupling
unification at a sufficiently high scale to account for proton
stability, and is automatically free from gauge anomalies.  In this
paper we focus on a $SO(10)$ GUT model in which the breaking to the SM
gauge group is driven by vevs of scalars in the $\fv_H \oplus
\bhts_H\oplus \ten_H$ representation. It has been recently shown that
this model is compatible with unification~\cite{Bertolini:2012im}, and 
that it can fit all charged fermion masses and mixings as well as the
low energy neutrino data~\cite{Dueck:2013gca,Joshipura:2011nn}, while
simultaneously  explaining the cosmological baryon asymmetry
via leptogenesis~\cite{Fong:2014gea}.
It is therefore interesting to see if this framework can also
accommodate automatically stable DM candidates in the fundamental
$\ten$ dimensional representation of the group.

\section{Motivations and general considerations}

$SO(10)$ is a rank five group and thus with respect to the SM model it
contains one additional Cartan generator that, upon breaking of the
unified group, can give rise to a new $U(1)$ gauge group factor.  The
$U(1)$ charges of the component fields are conventionally normalized
by setting the smallest charge equal to one.  Then, if $U(1)$ is
further broken by vevs of scalars carrying $n_1,n_2,\dots$ units of
charge with $N>1$ as their greatest common divisor, a discrete center
$Z_N \in U(1)$ remains unbroken~\cite{Ibanez:1991hv,Ibanez:1991pr,%
  DeMontigny:1993gy}.~\footnote{The fact that breaking $SO(10)$ with
  vevs in tensor representations can result in an unbroken $\z2$
  parity was already pointed out in~\cite{Kibble:1982ae}, in relation
  to the possible appearance of extended topological structures of
  cosmological relevance. We thank Q. Shafi for bringing this
  reference to our attention.}  
  In our setup $SO(10)$ is broken  to $U(1)_Q\times SU(3)_C$ by
vevs of scalars in $\mathbf{45}_H \oplus
\mathbf{\overline{126}}_H\oplus \mathbf{10}_H$, which are all $SO(10)$
tensor representations. With respect to the non-SM $U(1)$ factor
singled out in the maximal subgroup $SU(5)\times U(1)$, which is the
one whose breaking gives rise to the gauge discrete symmetry, all
$SO(10)$ tensor representations branch to $SU(5)\times U(1)$ fragments
which have even values of the $U(1)$ charge. The lowest charge value
for the fragments acquiring vevs is 2 [e.g.:  $\mathbf{10} \to
\mathbf{5}(2)\oplus \mathbf{\overline{5}}(-2)$] and therefore
a $\mathbb{Z}_2$ parity survives, which can guarantee the
stability of the lightest particles belonging to appropriately chosen
representations.

  Restricting to dimensions $D <
320$ we have that, depending if they are fermions or bosons, stable
particles appear in the following
representations~\cite{DeMontigny:1993gy,Kadastik:2009dj,%
  Kadastik:2009cu,Frigerio:2009wf}:
\begin{align}
\label{eq:even}
  &\text{Fermions:} && \ten, \fv, \irrep{54}, \irrep{120}, \hts, \irrep{210}, 
 \irrep{210'}\,,\\
\label{eq:odd}
 &\text{Bosons:} && \sixtn, \irrep{144} \,. 
\end{align}
For example, fermions in the vector $\mathbf{10}$ cannot decay 
into SM fermions in  the  $\mathbf{16}$ since this is a spinorial
representation for which all fragments under $SU(5)\times U(1)$ carry
odd $U(1)$  charges and  upon $U(1)$ breaking then 
acquire odd   $\mathbb{Z}_2$ parity.
Various proposals for $SO(10)$ DM candidates that are stabilized by
the $\z2$ parity of gauge origin have been put forth in the recent
literature: a dedicated analysis of scalar DM in the $\sixtn$ was
carried out in~\cite{Kadastik:2009dj,Kadastik:2009cu}, while the
possibility of fermionic DM in the $\mathbf{45}$ was addressed
in~\cite{Frigerio:2009wf} (see also~\cite{Arbelaez:2015ila} where the
$\mathbf{45}$ is allowed to mix with a $\ten$).  Other more general
studies regarding possible embeddings of DM in $SO(10)$ can be found
in~\cite{Mambrini:2015vna,Nagata:2015dma}.  Indeed, so far a special
attention has been devoted to DM in the scalar $\sixtn$ and in the
fermionic $\mathbf{45}$, and a possible reason for this might be the
fact that both these representations contain SM singlets. Needless to
say, identifying DM candidates with SM singlets can naturally explain
why all experimental direct detection (DD) searches have been eluded
so far.  In contrast, the $\ten$ dimensional vector representation of
$SO(10)$ has not attracted much attention, although this could well be
considered as the minimal choice.  Perhaps this is due to the fact
that the $\ten$ does not contain SM singlets, and in particular all
its states carry hypercharge, and thus couple to the $Z$ boson,
which might led to the conclusion that this possibility is excluded
by DD limits.\\

In this letter we argue that fermionic DM in the $\ten$ of $SO(10)$ is
instead a viable possibility.  Our main observation is that in a
scenario in which fermions in a vectorlike $\ten_\lL\oplus\ten_\lR $
acquire tree level masses via a Yukawa coupling with the
(antisymmetric) $\fv_H$, loop diagrams involving an insertion of
$W_{L}$-$W_{R}$ mixing produce a mass splitting between the two
lightest mass eigenstates, which (in our minimal realization) are two
neutral Dirac fermions.  We show that the neutral $Z_{L,R}$ gauge
bosons couple non-diagonally the light eigenstate to the heavier one
and, as a result, at the leading order only inelastic neutral current
scatterings of DM off target nuclei is allowed.  If the mass splitting
between the light and heavy mass eigenstates is larger than the
typical DM kinetic energy $E_K\sim 200\kev$, then the scattering is
kinematically forbidden and DD bounds are automatically evaded.  Since
the loop-induced splitting is suppressed by the RH gauge boson mass,
the previous requirement can be translated into an upper-bound on
$M^2_{W_R}$ which, combined with the lower bounds from flavour and CP
violating processes in the $K$ and $B$ meson
systems~\cite{Bertolini:2014sua} and from direct searches at the
LHC~\cite{ATLAS:2012ak,CMS:2012zv,Chatrchyan:2014koa}, results in
$2.9\tev \lsim M_{W_R} \lsim 25\tev$.  The fact that the null result
of DD DM searches constrains $M_{W_R}$ to lie at a relatively low
scale, can reinforce the hope that a rich phenomenology could be
within the reach of the LHC.

\bigskip

\section{The SO(10) framework}
\label{framework}

We assume that $SO(10)$ is broken at the unification scale to the
intermediate group $\mathcal{G}_I= SU(3)_C \otimes SU(2)_L \otimes
SU(2)_R \otimes U(1)_{B-L} $ by the vev of a $\fv_H$. $\mathcal{G}_I$
is then broken at an intermediate scale $\Lambda_I$ to the SM gauge
group $\mathcal{G}_{SM} =SU(3)_C \otimes SU(2)_L \otimes U(1)_Y$ by
the vev of $\bhts_H$, and finally $\mathcal{G}_{SM}$ is broken by
electroweak doublet vevs in a (complexified) $\ten_H$ down to
$SU(3)_c\times U(1)$:
\beqa{BC}
SO(10) &\stackrel{\vev{\fv_H}}{\longrightarrow}&  3_C  2_L 2_R 1_{B-L} \nn\\
&\stackrel{\vev{\bhts_H}
}{\longrightarrow}& 3_C  2_L 1_Y \otimes \z2 \nn\\
&\stackrel{\vev{\ten_H}}{\longrightarrow}& 3_C 1_Q \otimes \z2 \,.
\eeqa
In \eqn{BC} we have introduced for the gauge groups the short-hand
notation e.g.  $\mathcal{G}_I= 3_C 2_L 2_R 1_{B-L}$, and together with
the unbroken continuous gauge symmetries we have also written down the
discrete $\z2$ factor which survives down to the last breaking step.
Although the GUT symmetry breaking triggered by an adjoint $\fv_H$
together with a $\bhts_H$ develops instabilities at the tree level, it
has been recently shown that the inclusion of quantum corrections can
solve this problem and make the model
viable~\cite{Bertolini:2012im,Bertolini:2012az,Bertolini:2012be}.  The
first symmetry breaking is achieved via the $3_C2_L 2_R 1_{B-L}$
singlet contained in the $\vev{\fv_H}$.  The second step is driven by
a $\vev{\bhts_H}$ vev in the SM singlet direction which also provides
Majorana masses for the RH neutrinos, and the last step is driven by
the vevs of electroweak doublets contained in the $\ten_H$.  Note that
while a real $\ten_H$ is sufficient to drive the $\mathcal{G}_I\to
\mathcal{G}_{SM} $ symmetry breaking, a $\ten_H$ containing complex
fields is needed to reproduce realistic fermion
masses~\cite{Babu:1992ia,Bajc:2005zf} and in particular to accommodate
the $m_t/m_b$ mass ratio.  Moreover, to reproduce the complete charged
fermion mass spectrum accounting also for Yukawa non-unification of
the lepton and down-type quarks of the first two generations, a
contribution from the vevs of the electroweak doublets appearing in
the $\bhts_H$ is also necessary~\cite{Babu:1992ia}. These vevs are
unavoidably induced when $SU(2)_L\times U(1)_Y$ is broken by the
$\vev{\ten_H}$~\cite{Babu:1992ia,Bajc:2005zf}.  All in all, the masses
of the SM fermions are generated from the following Yukawa terms:
\beqn{stn} 
- {\cal L}_{\mathrm{SM}}= \sixtn_i\,\left( h_{ij}\ten_H +
  g_{ij}\ten^*_H + f_{ij}\bhts_H\right) \,\sixtn_j\,, 
\eeqn
where $h$, $g$ and $f$ are $3\times 3$ symmetric matrices in flavour
space and $i,j$ are family indices.  The fermion couplings to
$\ten^*_H$ can be forbidden by assigning to the fields a global $U(1)$
Peccei-Quinn
charge~\cite{Babu:1992ia,Bajc:2005zf,Altarelli:2013aqa}. In practice
this sets $g_{ij}\to 0$, simplifying the Yukawa structure of the
model, and providing a DM candidate for non-supersymmetric $SO(10)$
models in the form of axions.  In the spirit of avoiding the
introduction of additional symmetries, and given that we are
interested in a weakly interacting DM candidate, we will not follow
this route, and we allow for $g_{ij}\neq 0$.  Possible FCNC arising
from coupling quarks of the same type to two different Higgs doublets,
as it would happen in this situation, can be kept under control in
various ways e.g. by assuming a hierarchy $g\ll f$.
\\

\subsection*{Adding fermions in the vector representation}
\label{sec:DM} 

Let us now add to the $SO(10)$ model outlined above a pair of
fundamentals $\ten_\lL\oplus\ten_\lR$ containing new
fermions.\footnote{We denote $L$ and $R$ chiralities with calligraphic
  subscripts ($\lL,\lR$), while normal  subscripts
  ($\scriptstyle L, R$) label the $SU(2)$ gauge group factors.}  The
tensor product of two vectors of $SO(10)$ is:
\beqn{eq:tenten}
\ten \otimes \ten = \mathbf{1}^s \oplus \fv^a \oplus \mathbf{54}^s\,,
\eeqn
where the superscripts denote $s$ymmetric and $a$ntisymmetric
representations.
Although our model does not contain a $\mathbf{54}$ of fundamental
scalars, loop corrections can generate mass contributions that mimic
the coupling to (effective) representations, as long as these
couplings are allowed by the symmetries of the model.  To keep as
general as possible it is then convenient to write down all the
allowed gauge invariant Yukawa couplings, which are:
%
\beqa{tenRe} 
\nonumber 
\hspace{-1cm} - {\cal L}_{\mathrm{DM}}  &=&
\sum_{a=\lL,\lR} \ten_a \left(M_a+ \lambda_a\, \mathbf{54}_H\right) \ten_a 
\\
&+& \left[\ten_\lL \left(M +y\, \mathbf{45}_H + \lambda\, 
  \mathbf{54}_H\right) \ten_\lR + {\rm H.c.}\right]\,, 
\eeqa
where, in order not to over-clutter the expressions, we have left
understood the usual spinor notations.  It is instructive to
analyze these couplings in terms of representations of the
$SU(5) \subset SO(10)$. The branching rule for the $SO(10)$ vector is
$\ten = \sufv +\bsufv$ so that in the first line the invariant mass
term $M_{a}$ multiplies the $\sufv\cdot\bsufv$ singlet from the
product of the same $\ten_a$.  The second term involves the symmetric
$\mathbf{54} =\mathbf{15}+\mathbf{\overline{15}}+\mathbf{24}$ and,
besides containing a $\sufv\cdot\mathbf{24}\cdot\bsufv$ coupling
involving the $SU(5)$ adjoint, it also includes couplings of the
symmetric $\mathbf{15}$ to a pair of fundamentals:
$ \sufv\cdot\mathbf{\overline{15}}\cdot\sufv+{\rm c.c}$.  Let us note
at this point that if the colorless $SU(2)$ triplet contained in the
$\mathbf{\overline{15}}$ ($\mathbf{15}$) acquires a small vev, these
terms would generate a Majorana mass for the neutral components of the
fermion doublets in the $\sufv$ ($\bsufv$).  However, the same is not
true for the analogous term in the second line since it contains only
terms that couple two different $\ten$'s.  Finally, since the
$\mathbf{45}_H$ is antisymmetric, it must couple different
representations, and thus it appears only in the second line.

The model we will now study is specified by the following ingredients:
(i) the $\mathbf{54}_H$ is absent;  (ii) the adjoint vev $\vev{\fv_H}$
which can be written as:
\beqn{fvvev}
\vev{\fv_H}=\mathrm{diag}(a,a,a,b,b) \,\otimes\, 
\left(\begin{smallmatrix}
0&1 \cr
\!\text{-}1&0
\end{smallmatrix}\right)\,,
\eeqn
acquires a Dimopoulos-Wilczek structure~\cite{DW,Srednicki:1982aj}
with $a\sim \Lambda_{GUT} $ and $b/a\approx 0$ (since $b\neq 0$ breaks
$SU(2)_L \times SU(2)_R $ we require $b \lsim \Lambda_I$ in order to
respect the symmetry breaking pattern~\eqn{BC});
(iii) we set  $M_{a} \to 0$. This can be viewed as technically
natural since in this limit a global $U(1)$ symmetry
$\ten_{\lL,\lR}\to e^{i\alpha_{\lL,\lR}}\ten_{\lL,\lR}$ arises (we will briefly
comment below on the consequences of relaxing this assumption);  (iv)
finally, we will also work with $M \to 0$. 
This is just a simplification: a term proportional to $M$ 
preserves the  global $U(1)$ symmetry obtained for   
$\alpha_{\lL}=-\alpha_{\lR}$ which eventually
ensures the Dirac nature of the DM states (see next section), and as long as
$M \sim y\, b\ll a$ a non-vanishing $M$ would not change the analysis.
%

\section{Fermion spectrum and neutral current couplings}
\label{spectrum}

In the following we label
$SU(2)_L \otimes SU(2)_R \otimes SU(3)_C \otimes U(1)_{B-L}$
representations as $(d_L d_R d_C)_{B-L}$ where $d_{L,R,C}$ denote the
dimensions of the multiplets under the respective symmetry factor, and
$B-L$ gives the value of the $U(1)_{B-L}$ charge.

Each one of the two $\ten$'s contains one $(2_L 2_R 1_C)_0$ bi-doublet.
We  denote  the bi-doublet contained in $\ten_{\lL,\lR}$ as $\xi_{\lL,\lR}$,  
with  components: 
\beqn{bidoublet}
\xi_{\lL,\lR} = \begin{pmatrix}
\xi_{\lL,\lR}^{+-} & \xi_{\lL,\lR}^{++} \\
 \xi_{\lL,\lR}^{--} & \xi_{\lL,\lR}^{-+}
\end{pmatrix}\,, 
\eeqn 
where the superscripts carried by the component fields denote the
$SU(2)_L \otimes SU(2)_R$ isospin eigenvalues of $T_{3L,3R}$ in units
of $\pm \frac{1}{2}$. Given that the $\ten$ has vanishing
$U(1)_{B-L}$quantum number, the electromagnetic charge for this
representation is simply $ Q=T_{3L}+ T_{3R}$. Thus,
$\xi_{\lL,\lR}^{+-}$ and $\xi_{\lL,\lR}^{-+}$ are neutral fields,
while $\xi_{\lL,\lR}^{++},\,\xi_{\lL,\lR}^{--}$ have electric charge
$Q=\pm 1$.  The mass term for the neutral states arising from
$\vev{\fv_H}$ in \eq{fvvev} reads:
\beqn{eq:mb}
-\mathcal{L}_m = m_b\left[
\left(\xi_\lR^{+-}\right)^\dagger\, \xi_\lL^{+-}+
\left(\xi_\lR^{-+}\right)^\dagger\, \xi_\lL^{-+}
+ {\rm H.c.} \right]
\eeqn
with $ m_b = y\, b $.  A similar mass term can be written also for the
charged components $\xi_{\lL,\lR}^{--},\,\xi_{\lL,\lR}^{++}$ which at
lowest order are degenerate in mass with the neutral states.  However,
electromagnetic corrections from loops involving SM gauge bosons lift
this degeneracy inducing a charged-neutral mass difference
$m^\pm -m^0 \simeq 340\,$MeV~\cite{Cirelli:2005uq}, thus ensuring that the
lightest states are the neutral ones.\footnote{We have checked that
  loops of RH gauge bosons do not contribute. This is because
  the $\ten$ has vanishing $B-L$ charge, and $|T_{3R}|=\frac{1}{2}$ for
  both charged and neutral states, so that the corresponding loop
  contributions cancel in the mass difference. Therefore the SM
  result~\cite{Cirelli:2005uq} for the charged-neutral mass difference
  for fermion doublets holds also in this case.}

\Eq{eq:mb} describes a pair of neutral Dirac fermions
$\xi^{+-}\equiv \left(\xi^{+-}_\lL,\,\xi^{+-}_\lR\right)^T$ and
$\xi^{-+}\equiv \left(\xi^{-+}_\lL,\,\xi^{-+}_\lR\right)^T$ with
degenerate masses equal to $m_b$. However, these states get mixed via
loop diagrams (depicted in figure~\ref{fig:2}) involving two external
$\vev{\ten}_H$ vevs, which generate a mass  term: 
\beqn{eq:dm}  
-\mathcal{L}_\delta = \delta_m\left[
\left(\xi_\lR^{-+}\right)^\dagger\, \xi_\lL^{+-}+
\left(\xi_\lR^{+-}\right)^\dagger\, \xi_\lL^{-+}
+ {\rm H.c.} \right]\,.  
\eeqn
The tree level \eqn{eq:mb} and the loop induced mass term \eqn{eq:dm}
are sketched in figure~\ref{fig:1}.  Upon diagonalization of the mass
matrix (see the Appendix) we end up with one heavier ($\chi_h$) and
and one lighter ($\chi_l$) Dirac fermions, respectively with masses
\beqn{app-eigenvalues}
m_{h,l}       =  m_b \pm \delta_m \,. 
\eeqn
The important point is that the couplings of $\chi_{h,l}$ to the
neutral gauge bosons $Z_{L,R}$ are off-diagonal: both $Z_{L,R}$ couple
(with an opposite overall sign) to the vectorlike neutral current
\beqn{nc}
J_\mu^{nc}=\frac{1}{2}\;
\overline{\chi_{h}}   
\gamma_\mu \, \chi_{l}   + {\rm H.c.}.
\eeqn
Two ingredients are crucial for this result:
(i) a complex $\ten_H$: this is in any case needed to reproduce the SM
fermion mass pattern. The tensor product of two $\ten$'s contains in
its symmetric part the $\mathbf{54}$, which can be regarded here as an
effective $\mathbf{54}_H$ coupled to the fermions via the loop
diagram. Moreover, since $\ten_H$ develops vevs in the $L$-$R$ doublet
components, the effective $\mathbf{54}_H \subset \ten_H\otimes \ten_H$
contains a non-vanishing $SU(2)_L\times SU(2)_R$ breaking vev in
$(3_L,3_R,1_C)_0$, which provides an independent mass term to the
fermion bi-doublets.
(ii) a tree level mass coupling between $\ten_\lL$ and $\ten_\lR$: a
non-vanishing loop mass contribution can only appear if a chirality
flip can be inserted inside the loop diagram.  In our simplified setup
with $M_{\lL,\lR},M =0$ this can only be provided by $m_b$ (a
contribution from $M\neq 0$ which also couples $\ten_\lL$ to
$\ten_\lR$ would act in the same way with the replacement
$m_b \to m_b+M$).  Note that the global $U(1)$ symmetry mentioned at
the end of section~\ref{framework} remains unbroken also at the loop
level.\footnote{ Allowing for $M_{\lL,\lR}\neq 0$ would instead
  break this $U(1)$. As a result, we can expect that the two Dirac states
  will split into four Majorana fermions with a spectrum determined by
  the relative sizes of $m_b,\, \delta m$ and $M_{\lL,\lR}$.  }

\begin{figure}[t!]
\begin{center}
\includegraphics[width=.4\linewidth]{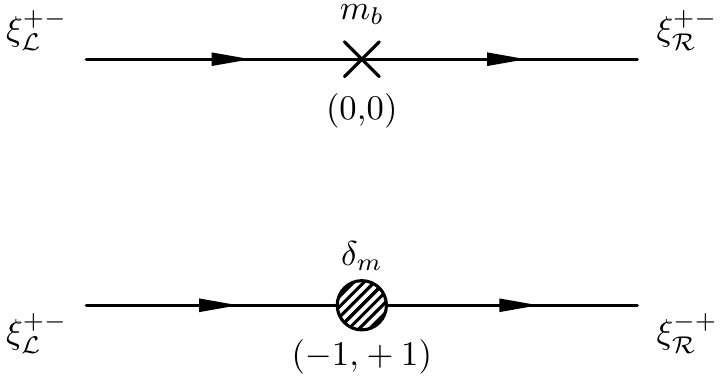}
 \caption{The tree level and the loop induced mass terms \eqns{eq:mb}{eq:dm}.
The numbers given in parenthesis refer to the  $T_3^{L,R}$ values carried 
by the mass insertions.} 
\label{fig:1}
 \end{center}
\end{figure}

\section{Constraints from direct detection}
\label{sec:dd}

The mass splitting $m_h-m_l=2\delta_m$ is an important quantity, since
for $m_{DM} \sim 1\,$TeV, the $Z_{L,R}$ mediated inelastic DM
scattering off target nuclei $\chi_l + N \to \chi_h + N$ is
kinematically forbidden only if
$2\delta_m \gsim 200\kev$~\cite{Nagata:2014aoa}, and only in this case
$\chi_l$ could have escaped DD DM searches. Let us then proceed to
estimate its value.
 
\begin{figure}[t!]
\begin{center}
  \includegraphics[width=.5\linewidth]{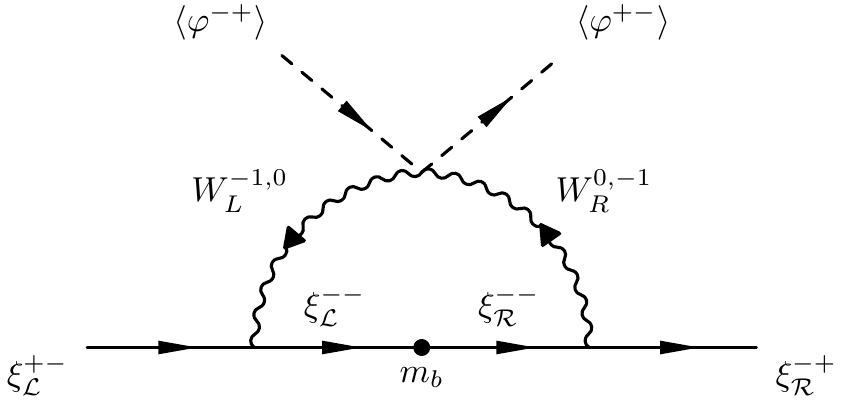}
  \caption{The one loop diagram that generates the mass term
    $\delta_{m}$ for 
    $\left(\xi_\lR^{-+}\right)^\dagger\, \xi_\lL^{+-}$
.  Inclusion of the $W_L\leftrightarrow W_R$ crossed
    diagram is left understood.  An analogous diagram contributes the
    same $\delta_m$ also for 
    $\left(\xi_\lR^{+-}\right)^\dagger\, \xi_\lL^{-+}$.}
  \label{fig:2}
\end{center}
\end{figure}

We denote the bi-doublet scalar contained in $\ten_H$ as $\phi$,
with components:
\beqn{bi10}
\phi = \begin{pmatrix}
\varphi^{+-} & \varphi^{++} \\
\varphi^{--} & \varphi^{-+} 
\end{pmatrix}\,, 
\eeqn
where the superscripts have the same meaning as for the fermions
\eqn{bidoublet}. Assuming for simplicity that there is only one scalar
bidoublet and that its vevs are real, we can write
\beqn{vevsbi}
\vev{\phi} = 
\begin{pmatrix}
v_u &  \\
 & v_d 
\end{pmatrix}\,, 
\eeqn
where $v_u^2+v_d^2 = v^2$ is the electroweak breaking vev.  One
diagram contributing to $\delta_m$ is depicted in figure~\ref{fig:2}.
The crossed diagrams should also be added, and a similar pair of
diagrams can be drawn for external fermions with exchanged LR isospin
labels ${\scriptstyle (+-)} \leftrightarrow {\scriptstyle (-+)}$.
Taking $M_{W_R}$ as the largest mass scale in the loop, the diagram
can be estimated as
\begin{align}\label{deltaM}
\frac{1}{2}\delta_m &\sim  
\frac{g^2_Lg^2_R}{16\pi^2}  
\frac {v_u v_d}{M^2_{W_R}}\,  m_b \nn\\
&\sim \frac{2\,\alpha}{4\pi s^2_W}  
\frac{v_u v_d}{v^2_R} \,  m_b = 5 \times 10^{-3} \;\vartheta_{LR}\, m_b\,, 
\end{align}
where we have used $M_{W_R} = g_R\, v_R/\sqrt{2}$ with $v_R$ the
$\mathcal{G}_I$ breaking vev of $\bhts_H$, $s_W= \sin\theta_W$ with
$\theta_W$ the Weinberg angle, and $m_b\gg \delta_m$ is to a very good
approximation the DM mass.  In the last expression we have introduced
the $W_R$-$W_L$ mixing parameter $\vartheta_{LR} = \frac{v_u
  v_d}{v^2_R}$ which is experimentally bounded in various ways.
Electroweak precision data set the upper limit $\vartheta_{LR}<
0.013 $~\cite{Langacker:1989xa,Czakon:1999ga}.  On the other hand,
$\vartheta_{LR}$ is bounded from above also by the ratio of gauge
boson masses squared:
\beqn{LRmix} 
\vartheta_{LR}= \frac{2 v_u v_d}{v^2_u+v^2_d} \frac{M^2_{W_L}}{M^2_{W_R}}\lsim
 \frac{M^2_{W_L}}{M^2_{W_R}}\,,
\eeqn 
where in the first equality we have approximated $g_R\approx g_L$.
Flavour and CP violating processes in the $K$ and $B$ meson systems
provide an absolute lower bound on the $SU(2)_R$ gauge bosons mass
$M_{W_R} > 2.9\,$TeV~\cite{Bertolini:2014sua}. Similar bounds on
$M_{W_R}$ have been also obtained from direct searches at the
LHC~\cite{ATLAS:2012ak,CMS:2012zv,Chatrchyan:2014koa}.  Using these
figures we obtain the conservative upper bound $\vartheta_{LR}
<7.7 \times 10^{-4}$.  The mass splitting $2\delta_m$ between the
light and heavy neutral states can then be bounded from above as:
\beqn{split}
2\delta_m \lsim 15 \>\left(\frac{2.9\tev}{M_{W_R}}\right)^2\,
\left(\frac{m_b}{1\tev}\right)\, \mev\,.
\eeqn
The constraint from DM non-observations in DD
experiments~\cite{Nagata:2014aoa} $2\delta_m \gsim 200\kev$ can then
be translated in the following {\em upper limit} on the $SU(2)_R$
gauge boson masses:
\beqn{MWR}
M_{W_R} \lsim 25 \left(\frac{m_b}{1\tev}\right)^{1/2} \tev \,.
\eeqn 
Indeed this result suggests the possibility of a non trivial interplay
between DM searches in DD experiments, and searches for new physics at
the LHC or at a future $100\,$TeV hadron collider.\footnote{Let us
  recall at this point that a certain number of anomalies have been
  recently reported by both {\sc ATLAS}~\cite{Aad:2015owa} and {\sc
    CMS}~\cite{Khachatryan:2014hpa,Khachatryan:2014gha} which could be
  explained by a low-scale L-R model with $M_{W_R}\sim 2 \tev$, see
e.g.,~\cite{Dobrescu:2015qna,Cheung:2015nha,Brehmer:2015cia,%
Gao:2015irw,Dobrescu:2015yba,Coloma:2015una,Deppisch:2015cua,%
Dev:2015pga,Bandyopadhyay:2015fka}.}

Before concluding this section, let us note that non-vanishing DD
 cross sections will appear at the loop level, with leading
 contributions involving the exchange of pairs of $SU(2)_L$ gauge
 bosons $(W_LW_L$ or $Z_LZ_L)$. The quantitative effects of the
 corresponding diagrams have been studied for example in
 refs.~\cite{Hisano:2011cs,Hill:2011be,Hill:2013hoa,Hill:2014yka}, and
 it was found that the resulting cross sections do not exceed $\sim
 \mathcal{O}(10^{-47})\;$cm$^2$, which is far below the current
 experimental bounds~\cite{Aprile:2012nq,Akerib:2013tjd}.  In the
 relevant mass range ($m_\chi \gsim \,$TeV) this remains also below
 the reach of next generation DD experiments~\cite{Cushman:2013zza}
 and close to the neutrino scattering background.

\section{Neutrino masses}
\label{neutrinos}
The relatively low value of the intermediate symmetry breaking scale
implied by \eqn{MWR} could be of some concern for what regards the
light neutrino masses. In general, the fact that the seesaw
mechanism~\cite{Minkowski:1977sc,Yanagida:1979as,Glashow:1979nm,%
  GellMann:1980vs,Mohapatra:1979ia} can be automatically embedded
within $SO(10)$ provides an elegant way to explain why the neutrino
masses are so suppressed.  $SO(10)$ unification implies relations
between the light neutrino, RH neutrinos, and up-type quark masses,
which generically require rather heavy RH neutrinos ($M_{N_R}\sim
m^2_u/m_\nu$). The natural range for $M_{N_R}$ is then loosely
determined by the up and top quark masses as $10^4\gev \lsim
M_{N_R}\lsim 10^{14}\gev$.  On the other hand $N_R$'s acquire their
masses from the same $\bhts_H$ vev that breaks $\mathcal{G}_I$ and
concurs to determine the value of $M_{W_R}$, and thus we would expect
their masses to be of the order of $M_{W_R}$, which remains bounded
from above by~\eqn{MWR}. Such a relatively low mass scale for the RH
neutrinos does not provide enough suppression.  One might then be
tempted to appeal to a different suppression mechanism.  For example
the inverse seesaw~\cite{Mohapatra:1986bd} is an elegant option that
allows to suppress neutrino masses even when $M_{N_R} \sim \tev$.
However, implementing the inverse seesaw requires the addition of a
$SO(10)$ singlet with a (small) Majorana mass term, which couples to
$N_R$ and to some scalar representation with non-vanishing vev.  Given
that $N_R \in \sixtn$ the only option for writing down a
renormalizable coupling is a scalar multiplet $\bsixtn_H$.  However,
$\vev{\bsixtn_H}\neq 0$ would break the $\z2$ parity thus allowing DM
decays.  A similar conclusion can be reached also in the case the new
fermion is not a $SO(10)$ singlet but is assigned to some suitable
$SO(10)$ representation.  We must then conclude that in our framework
the inverse seesaw does not provide a viable alternative to explain
the neutrino mass suppression.

A straightforward, although not so elegant, way out, is to appeal to
cancellations in the neutrino Dirac mass matrix. This relies on the
fact that, when projected onto SM multiplets, the fermion mass
matrices originating from \eqn{stn} acquire non trivial Clebsch-Gordan
coefficients that weight the various vev contributions and that are
different for different fermion species.  For the up-type quark and
Dirac neutrino masses we have~\cite{Fong:2014gea, Altarelli:2013aqa,
  Nath:2001uw}:
\beqa{fermions}
m_u &=& + (h v_u + g v_d) +\sqrt{3} f \kappa_u\,, \\
m_\nu^D &=& - (h v_u +g v_d) +3\sqrt{3} f \kappa_u\,, 
\eeqa
where $h,g,f$ are the symmetric Yukawa matrices introduced in
\eqn{stn} and $\kappa_u$ is the up-type induced doublet vev from
$\bhts_H$.  If we take e.g.  $M_{N_R} \sim 10\tev$, no particular
cancellation is needed for the mass entries related to the up quark
mass, while for those related to the heavy third generation, a tuning
in the cancellation of up to one part in $10^5$ is required. While
this is certainly unpleasant, we should not forget that
non-supersymmetric $SO(10)$ suffers a naturalness problem from the
start, which already requires a tuning in the theory at a much higher
level than $10^{-5}$.

 \begin{figure}[t!]
  \centering
  \includegraphics[width=.8\linewidth]{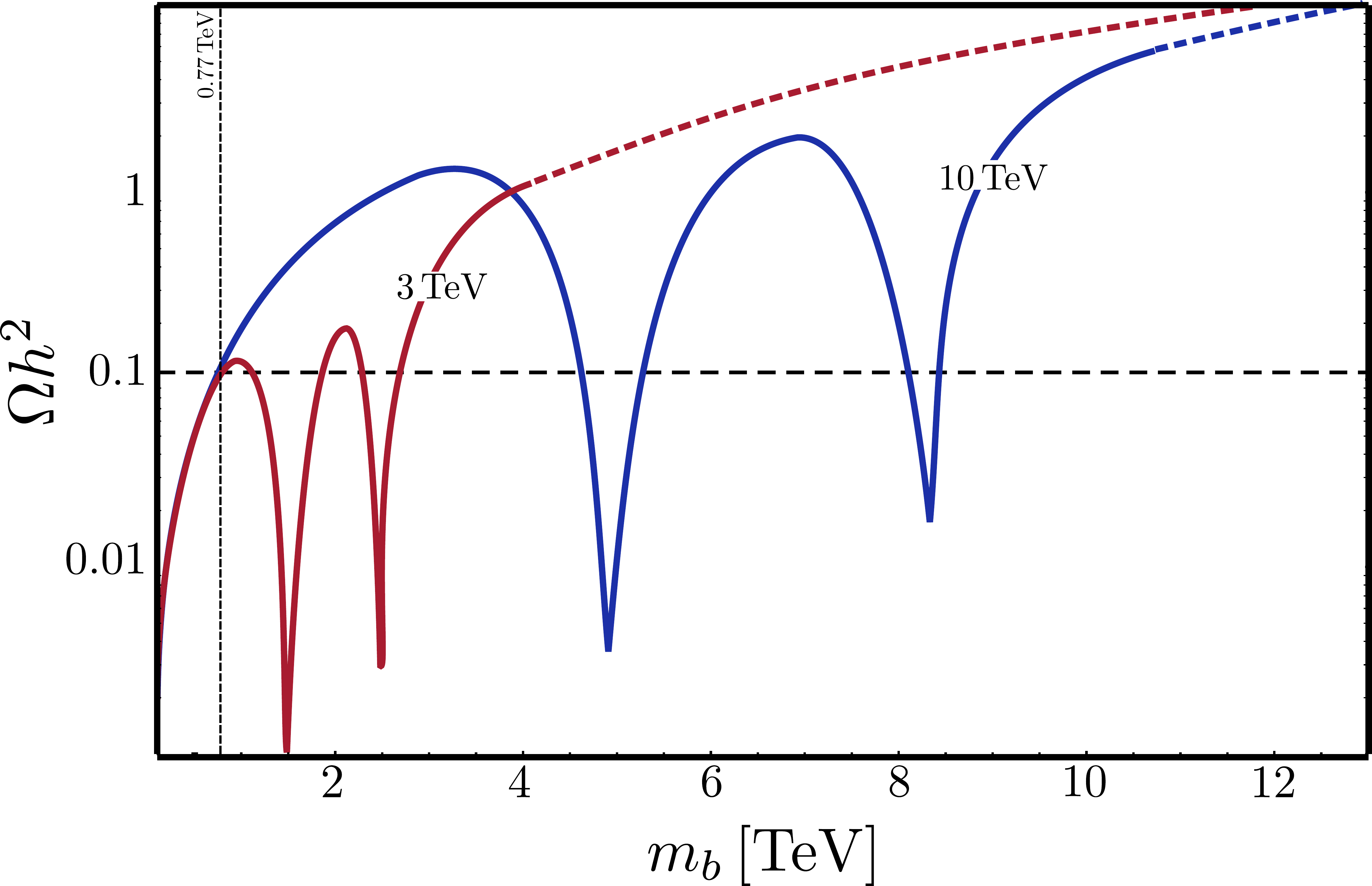}
  \caption{DM relic density as a function of $m_b$ for 
$M_{W_R}=3 \tev$ ($M_{Z_R}\sim 5\tev$) and 
$M_{W_R}=10 \tev$ ($M_{Z_R}\sim 17\tev$).}
  \label{fig:relic_density}
\end{figure}

\section{Relic density}
\label{relic}
We now turn to the calculation of the relic density of the DM
candidate $\chi_l$. Right above the intermediate scale (of order TeV)
our $SO(10)$ model corresponds essentially to a low-scale L-R
extension of the
SM~\cite{Mohapatra:1979ia,Senjanovic:1978ev,Senjanovic:1975rk,%
  Pati:1974yy,Mohapatra:1974gc,Mohapatra:1980yp} with the addition of
two (massless) fermion bi-doublets. After $SU(2)_R$ breaking, two
degenerate Dirac $SU(2)_L$ doublets with mass $m_b$ appear.  Below the
electroweak breaking scale the neutral members of these doublets
combine into two neutral fermions $\chi_h$ and $\chi_l$ with masses
$m_b \pm \delta_m$. Let us recall that $\delta m \sim
\mathcal{O}(\mev)$ while the mass difference with the heavier charged
partners is of about $340\mev$ (see the discussion in
section~\ref{spectrum})
so that both mass splittings are much smaller than the typical DM
freeze-out temperature, and have negligible effects on the
determination of the DM relic abundance. 
This is because in the limit of unbroken LR symmetry the fermion
states are degenerate members of gauge multiplets, and
`cohannihilation' channels can be simply taken into account by
including appropriate factors of gauge multiplicities both in the
annihilation process and in the counting of particle degrees of
freedom. However, in our numerical study we have kept track of the
small mass splittings induced by symmetry breaking thus
differentiating between annihilation and cohannihilation, but
the effects of this more refined treatment remain
irrelevant.  Of course, eventually the charged states and the heavier
neutral state $\chi_h$ will all decay to $\chi_l$, thus adding their
contribution to the DM relic abundance.
\\

Before tackling the calculation of the DM relic density in our model,
let us first recall the generic features of the analogous computation
in the case of a DM SM doublet.  The relic density of the neutral
component of an $SU(2)_L$ doublet of mass $m_{DM}>M_{Z_L}$ can
be cast as~\cite{Cirelli:2005uq}:
\beqn{relicMDM}
\Omega_{DM}\,h^2 \approx 0.1\, \frac{4.2\times 10^{-3} 
\tev^{-2}}{\vev{\sigma v}} \approx
		 0.1\, \left(\frac{m_{DM}}{1.1 \tev}\right)^2 \,,
\eeqn
where $\vev{\sigma v}$ is the thermally averaged annihilation
cross-section times the relative velocity, and includes SM gauge
annihilation and co-annihilation processes.  The scaling of
$\Omega_{DM}$ with the square of $m_{DM}$ follows from the scaling of
the annihilation cross section as $m_{DM}^{-2}$, when the DM mass is
larger than the SM gauge boson masses. Clearly, in the presence of two
quasi-degenerate doublets with mass $m_b$, the same value of
$\Omega_{DM}$ would be reproduced for $m_b = m_{DM}/\sqrt{2}$.
However, in our case in addition to the SM gauge interactions we have
new annihilation channels mediated by $Z_R$ and $W_R$.  For
$m_b\sim M_{W_R}$ the cross section does not scale simply as
$m_b^{-2}$, and in particular for $m_b \sim M_{W_R}/2$ and
$m_b \sim M_{Z_R}/2$ annihilation in the $s$-channel proceeds via
gauge boson resonances which drastically lowers the otherwise large
relic density implied by \eq{relicMDM}.  Therefore, in addition to the
solution at $0.77\, \tev$, which approximately holds for
$M_{Z_L} < m_b \ll M_{W_R}$ we expect other phenomenologically viable
values of $m_b$ which will depend on the values of the RH gauge boson
masses.

In order to explore quantitatively the mass parameter space for DM,
including also the effects of resonant annihilation and
co-annihilation via $Z_R,W_R$, we have carried out a numerical
analysis by implementing our model in {\sc
  Micromegas}~\cite{Belanger:2014vza}.\footnote{The model file was
  generated with {\sc Feynrules}~\cite{Alloul:2013bka} by modifying
  the model of Ref.~\cite{Roitgrund:2014zka}.}  The results are
summarized in figure~\ref{fig:relic_density} where the dependence of
the relic density on the DM mass $m_b$ is shown for two benchmark
values of $M_{W_R}$: $3 \tev$ and $10 \tev$ (the corresponding $Z_R$
mass values are respectively $5\tev$ and $17\tev$).
The first solution corresponds to the expected value
$m_b \simeq 0.77 \tev$. However, other values of $m_b$ become viable
close to the resonance regions $m_b \sim M_{W_R}/2$ and
$m_b  \sim M_{Z_R}/2$.  The width of these resonances is large
enough to allow for a clear separation of the two crossings that are
present for each resonance. All in all, we have five
phenomenologically viable values of $m_b$ for each value of
$M_{W_R}$. Eventually, for very large DM masses $m_b \gg M_{W_R}$ we recover
the scaling of \eq{relicMDM}. however, in this region the relic
densities are at least one
order of magnitude larger  than the observed $\Omega_{DM}$. \\

As we have seen in section~\ref{spectrum},
DM does not couple diagonally to the neutral $Z_{L,R}$ gauge bosons.
Then the leading annihilation channel for indirect detection searches
is into $W_LW_L$ and $Z_LZ_L$ (respectively via $t$-channel exchange
of $\chi^\pm$ and $\chi_h$) with comparable branching ratios for the
two diboson final states (for the DM solutions corresponding to
$m_b \sim M_{Z_R}/2$, see figure~\ref{fig:relic_density},
$W_LW_R$ final states are also kinematically accessible). The velocity-averaged 
cross-section for $\chi_l\bar\chi_l \to W_LW_L$ can be estimated as
$\langle \sigma_W |{\rm v}|\rangle \sim 
{\pi \alpha_g^2}/({32 m_l^2}) \sim 3\times 10^{-28} \left(
  {2\,{\rm TeV}}/{m_l}\right)^2 \,{\rm {cm}^3/s} $
with $\alpha_g$ the $SU(2)$ fine structure constant. A more accurate
estimate, including non-relativistic Sommerfeld corrections, gives for
the same mass range an enhancement up to one order of
magnitude~\cite{Cirelli:2007xd}. In spite of this the signal remains
well below the present limits
$\langle \sigma_W |{\rm v}|\rangle \lsim (10^{-25} - 10^{-24}) \,  
  {\rm cm^3/s} $
for the mass range $1\,{\rm TeV} < m_l < 4\,{\rm TeV}$~\cite{Ackermann:2015zua}.

As regards collider limits, at the LHC the most sensitive searches in our
scenario, in which $\chi_a=\chi_{l,h}, \chi^\pm$ are quasi-degenerate,
are monojet signatures from processes like $pp\to \chi_a \chi_b j$,
where the pair of $\chi$'s are produced via the $s$-channel exchange
of a SM gauge boson.  However, this signal is accompanied by large
backgrounds from $Z,W$+jets, which render the experimental search
particularly difficult. A dedicated analysis of signatures of
quasi-degenerate Higgsino like DM, which closely resembles our
scenario, indicates a surprisingly low reach $m_l \sim 250\,$GeV even
for LHC-13~\cite{Barducci:2015ffa}.  We have no reason to expect that
this limit could be largely exceeded in our case, so that we can
conclude that even the  non-resonant (lowest mass) DM solution with 
$m_l\sim 0.77\,$TeV remains unconstrained by collider searches.\\

Before concluding this section, one additional remark is in order. Our
Dirac DM candidates carry hypercharge $Y=T_{3R}$, which can be used to
distinguish particles (e.g. $\chi^{++},\chi^{-+}$) from antiparticles
(respectively $\chi^{--},\chi^{+-}$).  During their thermal history,
the $\chi$'s will unavoidably enter in chemical equilibrium with the
thermal bath, inheriting an asymmetry similar to that of all SM
hypercharged states. For example, in the temperature range $T_R>T>T_L$
($T_{R,L}$ denote the temperatures at which $SU(2)_{R,L}$ get broken)
that is the relevant range in which DM annihilates efficiently when
the DM mass is above a few TeV, processes like $\xi^{++} \xi^{-+}
\leftrightarrow W_R^{0,+1}$ which occur as long as $T\gsim M_{W_R}$,
or $2\leftrightarrow2$ scatterings mediated by $D=6$ effective
operators like $\frac{g_R^2}{M^2_{W_R}}\, (\bar \xi^{--} \gamma_\mu
\xi^{-+})\, (\bar u_R\gamma^\mu d_R)$ which occur when $T < M_{W_R}$,
will enforce chemical equilibrium between the $\chi$ system and the SM
thermal bath, and thus an asymmetry will develop in the DM sector as
well.  The issue whether such an asymmetry could play any relevant role in
determining the final DM relic density by quenching the annihilations
when the DM density becomes of the order of the density asymmetry was
recently studied in~\cite{Boucenna:2015haa} for the general case of
stable relics belonging to scalar and fermion hypercharged multiplets
of dimension $D\geq 2$.  It was found that for fermion doublets, as
for most of the other cases, the effects of the DM asymmetry on the
surviving relic density are generally negligible.  However, the
results of the analysis in~\cite{Boucenna:2015haa} rely on certain
assumptions, among which: (i) there are no new hypercharged particles
besides the multiplet in question; (ii) the same operator responsible
for the transfer of the asymmetry is also responsible for the mass
splitting between the neutral hypercharged states.

In the present case however, these two conditions are not satisfied:
(i) besides the new fermions in the bi-doublets, also the charged
gauge bosons $W_R$ carry hypercharge $Y=T_{3R}$; (ii) the loop
operator that induces the mass splitting (figure~\ref{fig:2}) is
generated only after $SU(2)_L$ breaking, while the asymmetry transfer
is mediated by tree level interactions with real or virtual $W_R$
bosons, and is most efficient well above $T_L$.  Moreover, resonantly
enhanced annihilation and co-annihilation, that were not present in
the scenario analyzed in~\cite{Boucenna:2015haa}, here play a very
important role, and many of the solutions for the correct DM relic density
are found in DM mass regions close to the gauge boson resonances.
Nevertheless, in spite of all these differences, given that
co-annihilation via $W_R$ exchange, that is one of the dominant
annihilation processes, does not suffer from any asymmetry-related
quenching, it is reasonable to expect that in most of the parameter
space the DM asymmetry will be largely uninfluential in determining
the final value of $\Omega_{DM}$.\footnote{The special cases in which
  the relic density is determined by annihilations close to the $Z_R$
  resonance could be an exception.}

\section{Conclusions}
Breaking the $SO(10)$ GUT group to the SM via the intermediate group
$SU(3)_C\otimes SU(2)_L\otimes SU(2)_R\otimes U(1)_{B-L} $ by means of
vacuum expectation values in $\fv_H \oplus\bhts_H\oplus \ten_H$
preserves an exact discrete gauge symmetry $\z2$, which ensures the
stability of the lightest among new fermions belonging to
$\ten$-dimensional vector representations. We have added to the
$SO(10)$ model two fermionic $\ten$'s, and we have argued that the
lightest stable states belonging to these representations correspond
to the four {\em neutral} members of the $SU(2)_L\otimes SU(2)_R$
bi-doublets contained in the two $\ten$'s. Thus, a first requirement
for viable DM candidates, namely electric charge neutrality, is
satisfied.  After $SU(2)_R$ breaking the four neutral states arrange
into two Dirac spinors describing two types of fermions that are
degenerate in mass, and after $SU(2)_L$ breaking loop corrections
involving the $W_R$ gauge bosons mix these states in such a way that:
(i) the mass eigenstates get splitted by an amount $\delta m$; (ii)
both the $Z_{L,R}$ neutral gauge bosons couple non-diagonally the
lighter neutral fermion to the heavier one.  Requiring that $\delta m$
is large enough to forbid neutral current inelastic scatterings of DM
into the heavier fermions allows to evade all the limits from DD DM
searches, in spite of the fact that DM is constituted by weakly
interacting Dirac fermions.  We have shown that the quantitative
requirement $\delta m \gsim 200\kev$ can be satisfied only if the
scale of $SU(2)_R$ breaking is not too large, and we have derived an
{\it upper} limit on the RH gauge bosons mass $M_{W_R} \lsim 25\,$TeV.
This upper bound strengthen the possibility that the LHC might detect
signatures of a low scale LR symmetry, and this is particularly
exciting in view of the anomalies recently reported by
ATLAS~\cite{Aad:2015owa} and
CMS~\cite{Khachatryan:2014hpa,Khachatryan:2014gha} which can all be
explained with  a RH gauge boson with mass $M_{W_R}\sim 2\tev$.\\

Before concluding, we should also point out some issues that within
our minimal scenario are left open, and that might deserve further
studies.  In section~\ref{neutrinos} we have pointed out that, as in
all $SO(10)$ derived low scale LR models, there is not enough
suppression for the light neutrino masses from the seesaw mechanism,
and to accommodate the neutrino mass scale we had to invoke some
amount of tuning in the Yukawa sector.  A related issue is the fact
that the scale of the RH neutrino masses is too low to allow for an
explanation of the cosmological baryon asymmetry via the standard
leptogenesis scenario~\cite{Fukugita:1986hr}
(see~\cite{Davidson:2008bu} for a review). This is because the RH
neutrinos are too light to provide sufficiently large CP violating
asymmetries~\cite{Davidson:2002qv}.  Finally, we have verified that
with the minimal particle content that we have assumed, gauge coupling
unification does not occur; however, we expect that it can be
recovered by adding new particles in suitable incomplete $SO(10)$
representations (see for example~\cite{Arbelaez:2013nga} for
different ways to recover gauge coupling unification in
low scale LR models).

\section*{Acknowledgments}
We thank Davide Meloni for discussions in the early stage of this
project.  We acknowledge financial support from the research grant
``Theoretical Astroparticle Physics" number 2012CPPYP7 under the
program PRIN 2012 funded by the Italian ``Ministero dell’Istruzione,
Universit\'a e della Ricerca'' (MIUR) and from the INFN ``Iniziativa
Specifica'' Theoretical Astroparticle Physics (TAsP-LNF).  SMB
acknowledges support of the Spanish MICINN's Consolider-Ingenio 2010
Programme under grant MultiDark CSD2009-00064 .

\vspace{1cm}  

\centerline{\bf \large Appendix}

 \appendix
\addtocontents{toc}{\protect\setcounter{tocdepth}{0}}
\renewcommand{\thesection}{\Alph{section}}

\section{Mass eigenstates and  gauge interactions}
\label{app:detailmass}

The two $(2_L 2_R 1_C)_0$ bi-doublet contained in the $\ten$ of $SO(10)$ 
can written in components as: 
\beqn{ap-bidoublet}
\xi_{\lL,\lR} = 
\begin{pmatrix}
\xi_{\lL,\lR}^{+-} & \xi_{\lL,\lR}^{++} \\
 \xi_{\lL,\lR}^{--} & \xi_{\lL,\lR}^{-+}
\end{pmatrix} \,.
\eeqn
$\xi_{\lL,\lR}$ are multiplets of Weyl fermions 
coupled to $SU(2)_L\times SU(2)_R$ gauge fields 
through the gauge-invariant kinetic Lagrangian:
\beqn{Weyl}
\mathcal{L}_K= 
\xi^\dagger_\lL\,i\overline{\sigma}_\mu D^\mu \xi_\lL + 
\xi^\dagger_\lR\, i{\sigma}_\mu D^\mu \xi_\lR\,,
\eeqn
where $\overline{\sigma}_\mu= (I,-\vec{\sigma})$ and   
$\sigma_\mu= (I,\vec{\sigma})$ with $\vec\sigma$
the Pauli matrices acting in Lorentz space, and 
\beqn{covariant}
D^\mu = \partial^\mu 
-i g_L \vec{W}^\mu_L \frac{\vec{\tau}_L}{2} 
-i g_R \vec{W}^\mu_R \frac{\lvec{\tau}_R}{2} \,, 
\eeqn
where $\vec{\tau}_L$  are Pauli matrices acting 
in $SU(2)_{L}$ group space, i.e. on the first
superscript labels of the multiplet components in 
\eqn{ap-bidoublet}, while  $\lvec{\tau}_R$ 
of $SU(2)_{R}$ act on the second labels, and  the reversed vector sign 
reminds that the action is from the right: 
$\xi \to \xi' = \vec{\tau}_L\,\xi \, \lvec{\tau}_R$.
Let us now define
\beqn{Lchiral}
\tilde\zeta_\lL = \sigma_2 \xi^*_\lR, \qquad
\tilde\zeta^\dagger_\lL = \xi^T_\lR\sigma_2\,. 
\eeqn
From the free Weyl equation for R-chirality spinors 
$i \sigma_\mu \partial^\mu \xi_\lR=0$ and using the relation
$\sigma_2\vec{\sigma}^*\sigma_2=-\vec{\sigma}$ it is easily seen that
$\tilde\zeta_\lL$ satisfies $i\overline{\sigma}_\mu\partial^\mu\tilde\zeta_\lL=0$ 
and thus it contains $L$-chirality spinors. In terms of $\tilde\zeta_\lL$ 
the second term in \eqn{Weyl} can be rewritten as:
\beqa{zetaL} \nonumber \hspace{-.6cm} \xi^\dagger_\lR\, i{\sigma}_\mu
D^\mu \xi_\lR \!\!\!&=&\!\!\!  \tilde\zeta^\dagger_\lL\,
i\overline{\sigma}_\mu\,
\partial^\mu  \tilde\zeta_\lL \\  
\!\!\!&+&\!\!\! \tilde\zeta^\dagger_\lL i\overline{\sigma}_\mu\left[
i g_L \vec{W}^\mu_L \frac{\vec{\tau\,}^T_L}{2} 
\right.+\left.i g_R \vec{W}^\mu_R \frac{\lvec{\tau}^T_R}{2}
\right]\!\!\tilde\zeta_\lL\,\;\ \   
\eeqa

where we have integrated by parts the derivative term (neglecting a
$4$-divergence) and an overall change of sign is due to 
anticommutation of the fermion fields.  \Eqn{zetaL} shows
explicitly that $\zeta_\lL$ transforms in the
$SU(2)_L\otimes SU(2)_R$ conjugate representation
$ (\mathbf{\overline{2}}, \mathbf{\overline{2}}) $ with generators
$\overline{\vec\tau}= - \vec\tau^*= \tau_2\vec\tau\,\tau_2$, where the
last relation expresses the pseudoreality of $SU(2)$
representations. It is then convenient to define new $\zeta_\lL$
multiplets transforming similarly to $\xi_\lL$ in
$(\mathbf{2}, \mathbf{2})$:
\beqa{newzeta}
\nonumber
\zeta_\lL &=& \begin{pmatrix}
\zeta_\lL^{+-} & \zeta_\lL^{++} \cr
\zeta_\lL^{--} & \zeta_\lL^{-+}
\end{pmatrix} 
= \tau_{2L}\,\tilde\zeta_\lL\,\tau_{2R} \\
&=&
\begin{pmatrix}
\sigma_2 \left(\xi_\lR^{-+}\right)^* & -\sigma_2 \left(\xi_\lR^{--}\right)^*   \cr
-\sigma_2 \left(\xi_\lR^{++}\right)^* & \sigma_2\left(\xi_\lR^{+-}\right)^*
\end{pmatrix}   \,. 
\eeqa
Focusing now on the neutral fermions, let us define:
\beqn{Psi}
\Psi_1 = 
\begin{pmatrix}
 \xi_\lL^{+-}\cr 
\xi_\lL^{-+}  
\end{pmatrix}, \qquad 
\Psi_2 = 
\begin{pmatrix}
 \zeta_\lL^{+-} \cr 
  \zeta_\lL^{-+}
\end{pmatrix}\,. 
\eeqn
The  neutral current interactions for $\Psi_i$ ($i=1,2$) 
read:
\beqa{neutral}
\nonumber
\mathcal{L}^{nc} &=&
J_\mu^{nc} \cdot \left(g_LW^\mu_{3L} - g_R W^\mu_{3R}\right)\,, \\
J_\mu^{nc} &=& - \sum_{i=1,2} \Psi_i^\dagger\, i\overline{\sigma}_\mu 
\, T_3 \, 
\Psi_i\,, 
\eeqa
with $T_3 = {\rm diag}\left(+\frac{1}{2},\,-\frac{1}{2}\right)$. In
order to write $J_\mu^{nc}$ in terms of the mass eigenstates, let us
study how the mass terms are rewritten from the original basis of
LH and RH Weyl spinors $\xi_{\lL,\lR}$ to our basis of
LH spinors $\xi_\lL$ and $\zeta_\lL$.  The fermion bilinears
multiplying the tree level mass term for charged and neutral
states~\eqn{eq:mb}
are rewritten as:
\beqa{ap-mass1}
\hspace{-0.7cm}
m_b {\rm Tr}\left(\xi^\dagger_\lR\xi_\lL + \xi^\dagger_\lL\xi_\lR \right)  
\!\!\!&=&\!\!\! 
-m_b{\rm Tr} \left(
\zeta^T_\lL\sigma_2 \xi_\lL +
\xi^\dagger_\lL\sigma_2 \zeta^*_\lL
\right)  \,.
\eeqa
Written explicitly:
\beqa{ap-mass2} 
\nonumber \hspace{-0.5cm} {\rm Tr} \left(
  \zeta^T_\lL\sigma_2 \xi_\lL\right) \!\!\!&=&\!\!\! -
\left(\zeta^{+-}_\lL\right)^T\sigma_2\xi^{-+}_\lL -
\left(\zeta^{-+}_\lL\right)^T\sigma_2\xi^{+-}_\lL  \\
\!\!\!&\ &\!\!\! + \left(\zeta^{--}_\lL\right)^T\sigma_2\xi^{++}_\lL +
\left(\zeta^{++}_\lL\right)^T\sigma_2\xi^{--}_\lL \,.
\eeqa 
Similarly, the $SU(2)_L\otimes SU(2)_R$ breaking loop induced mass
term for the neutral states~\eqn{eq:dm} reads:
\beqa{ap-loopmass}
\delta_m \left[\left(\zeta^{+-}_\lL\right)^T\sigma_2\,\xi^{+-}_\lL  + 
\left(\zeta^{-+}_\lL\right)^T\sigma_2\,\xi^{-+}_\lL  \right]\,,
\eeqa
so that the Lagrangian for the masses of the neutral states is:
\beqn{neut-mass}
-\mathcal{L}_m^0 = \Psi_2^T \sigma_2\,\mathcal{M}\, \Psi_1 + {\rm H.c.} 
\eeqn
with 
\beqn{calM}
\mathcal{M}=
\begin{pmatrix}
\delta & -m_b \cr 
- m_b & \delta  
\end{pmatrix} \,. 
\eeqn
Being $\mathcal{M}$ symmetric it can be factorized as
$\mathcal{M}= V^T\, M\, V$ with $V$ unitary and $M$ diagonal with real
and positive eigenvalues.  The two eigenvalues are
$m_{h,l}=m_b\pm \delta$, and the heavy and light mass eigenstates
$\chi_{1}=(\chi_{1h},\,\chi_{1l})^T$ and
$\chi_{2}=(\chi_{2h},\,\chi_{2l})^T$ are given by
$\chi_{1,2} = V \Psi_{1,2}$ with
\beqn{V}
V = \frac{1}{\sqrt{2}}\begin{pmatrix}
-1& 1\cr
i& i
\end{pmatrix} \,.
\eeqn
By redefining now $\chi_{1}=\chi_{\lL}$ and
$ \chi_\lR=-\sigma_2\chi_2^* $, 
the mass term can be written as:
\beqn{newmass}
-\mathcal{L}_m^0 = 
\overline{\chi_l} \, m_l\, \chi_l +
\overline{\chi_h} \, m_h\, \chi_h\,,
\eeqn
where we have introduced the four spinor $\chi_l =
\left[(\chi_l)_{\lL},\, (\chi_l)_{\lR}\right]^T$ and a similar one for
$\chi_h$, and we have adopted the usual convention
$\overline{\chi_l}=\chi_l^\dagger \gamma_0$ with $\gamma_0 =
\left(\begin{smallmatrix}0&I\cr I&0\end{smallmatrix}\right)$ in the
chiral basis.  Thus, upon diagonalization of the mass matrix the new
fermions organize into two Dirac mass eigenstates splitted in mass by
$2\delta $.

Coming back to the neutral current gauge interactions, 
after rotating the interaction eigenstates $\Psi_i$  in \eqn{neutral} 
onto  mass eigenstates, the neutral current reads:
\beqa{neutralmass}
\nonumber 
\hspace{-0.3cm}
J_\mu^{nc} 
 \!\!\!&=& \!\!\!
- \sum_{i=1,2}\!\! \chi^\dagger_i i\overline{\sigma}_\mu  
\left( V\,T_3 V^\dagger\right) \chi_i  \\ 
&=& 
\frac{1}{2}\;
\overline{\chi_{h}}   
\gamma_\mu \, \chi_{l}   + {\rm H.c.}
\,, 
\eeqa
with
$\gamma_0\gamma_\mu = \left(\begin{smallmatrix}\overline{\sigma}_\mu &0\cr
   0& \sigma_\mu \end{smallmatrix}\right)$.
We see that the neutral gauge bosons couple the light mass eigenstates
to the heavy ones, a result that follows from the fact that
$V\,T_3 V^\dagger = -\frac{1}{2} \tau_2$ is anti-diagonal ($ \tau_2$
denotes the second Pauli matrix, but with no relation here with gauge group
factors or spinors).  Obviously, since
$V \partial_\mu V^\dagger = \partial_\mu$ the purely kinetic term for
the mass eigenstates remains diagonal.

 \bibliographystyle{modified}


\begin{thebibliography}{10}

\bibitem{Servant:2002aq}
G.~Servant and T.~M.~P. Tait, {\it {Is the lightest Kaluza-Klein particle a
  viable dark matter candidate?}},  {\em Nucl. Phys.} {\bf B650} (2003)
  391--419 [\href{http://arXiv.org/abs/hep-ph/0206071}{{\tt
  arXiv:hep-ph/0206071}}].

\bibitem{Cheng:2004yc}
H.-C. Cheng and I.~Low, {\it {Little hierarchy, little Higgses, and a little
  symmetry}},  {\em JHEP} {\bf 08} (2004) 061
  [\href{http://arXiv.org/abs/hep-ph/0405243}{{\tt arXiv:hep-ph/0405243}}].

\bibitem{Ma:2006km}
E.~Ma, {\it {Verifiable radiative seesaw mechanism of neutrino mass and dark
  matter}},  {\em Phys. Rev.} {\bf D73} (2006) 077301
  [\href{http://arXiv.org/abs/hep-ph/0601225}{{\tt arXiv:hep-ph/0601225}}].

\bibitem{Farzan:2012sa}
Y.~Farzan and E.~Ma, {\it {Dirac neutrino mass generation from dark matter}},
  {\em Phys. Rev.} {\bf D86} (2012) 033007
  [\href{http://arXiv.org/abs/1204.4890}{{\tt arXiv:1204.4890}}].

\bibitem{Deshpande:1977rw}
N.~G. Deshpande and E.~Ma, {\it {Pattern of Symmetry Breaking with Two Higgs
  Doublets}},  {\em Phys. Rev.} {\bf D18} (1978) 2574.

\bibitem{Barbieri:2006dq}
R.~Barbieri, L.~J. Hall and V.~S. Rychkov, {\it {Improved naturalness with a
  heavy Higgs: An Alternative road to LHC physics}},  {\em Phys. Rev.} {\bf
  D74} (2006) 015007 [\href{http://arXiv.org/abs/hep-ph/0603188}{{\tt
  arXiv:hep-ph/0603188}}].

\bibitem{LopezHonorez:2006gr}
L.~Lopez~Honorez, E.~Nezri, J.~F. Oliver and M.~H.~G. Tytgat, {\it {The Inert
  Doublet Model: An Archetype for Dark Matter}},  {\em JCAP} {\bf 0702} (2007)
  028 [\href{http://arXiv.org/abs/hep-ph/0612275}{{\tt arXiv:hep-ph/0612275}}].

\bibitem{Krauss:1988zc}
L.~M. Krauss and F.~Wilczek, {\it {Discrete Gauge Symmetry in Continuum
  Theories}},  {\em Phys. Rev. Lett.} {\bf 62} (1989) 1221.

\bibitem{Hamaguchi:1998wm}
K.~Hamaguchi, Y.~Nomura and T.~Yanagida, {\it {Superheavy dark matter with
  discrete gauge symmetries}},  {\em Phys. Rev.} {\bf D58} (1998) 103503
  [\href{http://arXiv.org/abs/hep-ph/9805346}{{\tt arXiv:hep-ph/9805346}}].

\bibitem{Hamaguchi:1998nj}
K.~Hamaguchi, Y.~Nomura and T.~Yanagida, {\it {Longlived superheavy dark matter
  with discrete gauge symmetries}},  {\em Phys. Rev.} {\bf D59} (1999) 063507
  [\href{http://arXiv.org/abs/hep-ph/9809426}{{\tt arXiv:hep-ph/9809426}}].

\bibitem{Martin:1992mq}
S.~P. Martin, {\it {Some simple criteria for gauged R-parity}},  {\em Phys.
  Rev.} {\bf D46} (1992) 2769--2772
  [\href{http://arXiv.org/abs/hep-ph/9207218}{{\tt arXiv:hep-ph/9207218}}].

\bibitem{Batell:2010bp}
B.~Batell, {\it {Dark Discrete Gauge Symmetries}},  {\em Phys. Rev.} {\bf D83}
  (2011) 035006 [\href{http://arXiv.org/abs/1007.0045}{{\tt arXiv:1007.0045}}].

\bibitem{Banks:1989zw}
T.~Banks, {\it {Report on Progress in Wormhole Physics}},  {\em Physicalia
  Mag.} {\bf 12} (1990) 19--68.

\bibitem{Giddings:1988cx}
S.~B. Giddings and A.~Strominger, {\it {Loss of Incoherence and Determination
  of Coupling Constants in Quantum Gravity}},  {\em Nucl. Phys.} {\bf B307}
  (1988) 854.

\bibitem{Coleman:1988tj}
S.~R. Coleman, {\it {Why There Is Nothing Rather Than Something: A Theory of
  the Cosmological Constant}},  {\em Nucl. Phys.} {\bf B310} (1988) 643.

\bibitem{Gilbert:1989nq}
G.~Gilbert, {\it {Wormhole induced proton decay}},  {\em Nucl. Phys.} {\bf
  B328} (1989) 159.

\bibitem{Mambrini:2015sia}
Y.~Mambrini, S.~Profumo and F.~S. Queiroz, {\it {Dark Matter and Global
  Symmetries}},  \href{http://arXiv.org/abs/1508.06635}{{\tt
  arXiv:1508.06635}}.

\bibitem{Boucenna:2012rc}
M.~S. Boucenna, R.~A. Lineros and J.~W.~F. Valle, {\it {Planck-scale effects on
  WIMP dark matter}},  {\em Front. Phys.} {\bf 1} (2013) 34
  [\href{http://arXiv.org/abs/1204.2576}{{\tt arXiv:1204.2576}}].

\bibitem{Minkowski:1977sc}
P.~Minkowski, {\it {mu $\to$ e gamma at a Rate of One Out of 1-Billion Muon
  Decays?}},  {\em Phys.Lett.} {\bf B67} (1977) 421.

\bibitem{Yanagida:1979as}
T.~Yanagida, {\it {Horizontal symmetry and masses of neutrinos}},  {\em
  Conf.Proc.} {\bf C7902131} (1979) 95--99.

\bibitem{Glashow:1979nm}
S.~L. Glashow, {\it {The Future of Elementary Particle Physics}},  {\em NATO
  Sci. Ser. B} {\bf 61} (1980) 687.

\bibitem{GellMann:1980vs}
M.~Gell-Mann, P.~Ramond and R.~Slansky, {\it {Complex Spinors and Unified
  Theories}},  {\em Conf.Proc.} {\bf C790927} (1979) 315--321
  [\href{http://arXiv.org/abs/1306.4669}{{\tt arXiv:1306.4669}}].

\bibitem{Mohapatra:1979ia}
R.~N. Mohapatra and G.~Senjanovic, {\it {Neutrino Mass and Spontaneous Parity
  Violation}},  {\em Phys.Rev.Lett.} {\bf 44} (1980) 912.

\bibitem{Bertolini:2012im}
S.~Bertolini, L.~Di~Luzio and M.~Malinsky, {\it {Seesaw Scale in the Minimal
  Renormalizable SO(10) Grand Unification}},  {\em Phys. Rev.} {\bf D85} (2012)
  095014 [\href{http://arXiv.org/abs/1202.0807}{{\tt arXiv:1202.0807}}].

\bibitem{Dueck:2013gca}
A.~Dueck and W.~Rodejohann, {\it {Fits to SO(10) Grand Unified Models}},  {\em
  JHEP} {\bf 09} (2013) 024 [\href{http://arXiv.org/abs/1306.4468}{{\tt
  arXiv:1306.4468}}].

\bibitem{Joshipura:2011nn}
A.~S. Joshipura and K.~M. Patel, {\it {Fermion Masses in SO(10) Models}},  {\em
  Phys. Rev.} {\bf D83} (2011) 095002
  [\href{http://arXiv.org/abs/1102.5148}{{\tt arXiv:1102.5148}}].

\bibitem{Fong:2014gea}
C.~S. Fong, D.~Meloni, A.~Meroni and E.~Nardi, {\it {Leptogenesis in SO(10)}},
  {\em JHEP} {\bf 01} (2015) 111 [\href{http://arXiv.org/abs/1412.4776}{{\tt
  arXiv:1412.4776}}].

\bibitem{Ibanez:1991hv}
L.~E. Ibanez and G.~G. Ross, {\it {Discrete gauge symmetry anomalies}},  {\em
  Phys. Lett.} {\bf B260} (1991) 291--295.

\bibitem{Ibanez:1991pr}
L.~E. Ibanez and G.~G. Ross, {\it {Discrete gauge symmetries and the origin of
  baryon and lepton number conservation in supersymmetric versions of the
  standard model}},  {\em Nucl. Phys.} {\bf B368} (1992) 3--37.

\bibitem{DeMontigny:1993gy}
M.~De~Montigny and M.~Masip, {\it {Discrete gauge symmetries in supersymmetric
  grand unified models}},  {\em Phys. Rev.} {\bf D49} (1994) 3734--3740
  [\href{http://arXiv.org/abs/hep-ph/9309312}{{\tt arXiv:hep-ph/9309312}}].

\bibitem{Kibble:1982ae}
T.~W.~B. Kibble, G.~Lazarides and Q.~Shafi, {\it {Strings in SO(10)}},  {\em
  Phys. Lett.} {\bf B113} (1982) 237.

\bibitem{Kadastik:2009dj}
M.~Kadastik, K.~Kannike and M.~Raidal, {\it {Matter parity as the origin of
  scalar Dark Matter}},  {\em Phys. Rev.} {\bf D81} (2010) 015002
  [\href{http://arXiv.org/abs/0903.2475}{{\tt arXiv:0903.2475}}].

\bibitem{Kadastik:2009cu}
M.~Kadastik, K.~Kannike and M.~Raidal, {\it {Dark Matter as the signal of Grand
  Unification}},  {\em Phys. Rev.} {\bf D80} (2009) 085020
  [\href{http://arXiv.org/abs/0907.1894}{{\tt arXiv:0907.1894}}]. [Erratum:
  Phys. Rev.D81,029903(2010)].

\bibitem{Frigerio:2009wf}
M.~Frigerio and T.~Hambye, {\it {Dark matter stability and unification without
  supersymmetry}},  {\em Phys. Rev.} {\bf D81} (2010) 075002
  [\href{http://arXiv.org/abs/0912.1545}{{\tt arXiv:0912.1545}}].

\bibitem{Arbelaez:2015ila}
C.~Arbelaez, R.~Longas, D.~Restrepo and O.~Zapata, {\it {Fermion dark matter
  from SO(10)}},  \href{http://arXiv.org/abs/1509.06313}{{\tt
  arXiv:1509.06313}}.

\bibitem{Mambrini:2015vna}
Y.~Mambrini, N.~Nagata, K.~A. Olive, J.~Quevillon and J.~Zheng, {\it {Dark
  matter and gauge coupling unification in nonsupersymmetric SO(10) grand
  unified models}},  {\em Phys. Rev.} {\bf D91} (2015), no.~9 095010
  [\href{http://arXiv.org/abs/1502.06929}{{\tt arXiv:1502.06929}}].

\bibitem{Nagata:2015dma}
N.~Nagata, K.~A. Olive and J.~Zheng, {\it {Weakly-Interacting Massive Particles
  in Non-supersymmetric SO(10) Grand Unified Models}},
  \href{http://arXiv.org/abs/1509.00809}{{\tt arXiv:1509.00809}}.

\bibitem{Bertolini:2014sua}
S.~Bertolini, A.~Maiezza and F.~Nesti, {\it {Present and Future K and B Meson
  Mixing Constraints on TeV Scale Left-Right Symmetry}},  {\em Phys. Rev.} {\bf
  D89} (2014), no.~9 095028 [\href{http://arXiv.org/abs/1403.7112}{{\tt
  arXiv:1403.7112}}].

\bibitem{ATLAS:2012ak}
{\bf ATLAS}, G.~Aad {\em et.~al.}, {\it {Search for heavy neutrinos and
  right-handed $W$ bosons in events with two leptons and jets in $pp$
  collisions at $\sqrt{s}=7$ TeV with the ATLAS detector}},  {\em Eur. Phys.
  J.} {\bf C72} (2012) 2056 [\href{http://arXiv.org/abs/1203.5420}{{\tt
  arXiv:1203.5420}}].

\bibitem{CMS:2012zv}
{\bf CMS}, S.~Chatrchyan {\em et.~al.}, {\it {Search for heavy neutrinos and
  W[R] bosons with right-handed couplings in a left-right symmetric model in pp
  collisions at sqrt(s) = 7 TeV}},  {\em Phys. Rev. Lett.} {\bf 109} (2012)
  261802 [\href{http://arXiv.org/abs/1210.2402}{{\tt arXiv:1210.2402}}].

\bibitem{Chatrchyan:2014koa}
{\bf CMS}, S.~Chatrchyan {\em et.~al.}, {\it {Search for W' $\to $ tb decays in
  the lepton + jets final state in pp collisions at $\sqrt{s}$ = 8 TeV}},  {\em
  JHEP} {\bf 05} (2014) 108 [\href{http://arXiv.org/abs/1402.2176}{{\tt
  arXiv:1402.2176}}].

\bibitem{Bertolini:2012az}
S.~Bertolini, L.~Di~Luzio and M.~Malinsky, {\it {Towards a New Minimal SO(10)
  Unification}},  {\em AIP Conf. Proc.} {\bf 1467} (2012) 37--44
  [\href{http://arXiv.org/abs/1205.5637}{{\tt arXiv:1205.5637}}].

\bibitem{Bertolini:2012be}
M.~Malinsky, S.~Bertolini and L.~Di~Luzio, {\it {Structure and prospects of the
  simplest SO(10) GUTs}},  \href{http://arXiv.org/abs/1210.3789}{{\tt
  arXiv:1210.3789}}. [AIP Conf. Proc.1534,293(2012)].

\bibitem{Babu:1992ia}
K.~S. Babu and R.~N. Mohapatra, {\it {Predictive neutrino spectrum in minimal
  SO(10) grand unification}},  {\em Phys. Rev. Lett.} {\bf 70} (1993)
  2845--2848 [\href{http://arXiv.org/abs/hep-ph/9209215}{{\tt
  arXiv:hep-ph/9209215}}].

\bibitem{Bajc:2005zf}
B.~Bajc, A.~Melfo, G.~Senjanovic and F.~Vissani, {\it {Yukawa sector in
  non-supersymmetric renormalizable SO(10)}},  {\em Phys. Rev.} {\bf D73}
  (2006) 055001 [\href{http://arXiv.org/abs/hep-ph/0510139}{{\tt
  arXiv:hep-ph/0510139}}].

\bibitem{Altarelli:2013aqa}
G.~Altarelli and D.~Meloni, {\it {A non supersymmetric SO(10) grand unified
  model for all the physics below $M_{GUT}$}},  {\em JHEP} {\bf 1308} (2013)
  021 [\href{http://arXiv.org/abs/1305.1001}{{\tt arXiv:1305.1001}}].

\bibitem{DW}
S.~Dimopoulos and F.~Wilczek {\em The Unity of the Fundamental Interactions,
  Proceedings of the 19th Course of the International School of Subnuclear
  Physics, Erice, Italy (1981). Edited by A. Zichichi (Plenum Press, New York,
  1983)}.

\bibitem{Srednicki:1982aj}
M.~Srednicki, {\it {Supersymmetric Grand Unified Theories and the Early
  Universe}},  {\em Nucl. Phys.} {\bf B202} (1982) 327.

\bibitem{Cirelli:2005uq}
M.~Cirelli, N.~Fornengo and A.~Strumia, {\it {Minimal dark matter}},  {\em
  Nucl.Phys.} {\bf B753} (2006) 178--194
  [\href{http://arXiv.org/abs/hep-ph/0512090}{{\tt arXiv:hep-ph/0512090}}].

\bibitem{Nagata:2014aoa}
N.~Nagata and S.~Shirai, {\it {Electroweakly-Interacting Dirac Dark Matter}},
  {\em Phys. Rev.} {\bf D91} (2015), no.~5 055035
  [\href{http://arXiv.org/abs/1411.0752}{{\tt arXiv:1411.0752}}].

\bibitem{Langacker:1989xa}
P.~Langacker and S.~U. Sankar, {\it {Bounds on the Mass of W(R) and the
  W(L)-W(R) Mixing Angle xi in General SU(2)-L x SU(2)-R x U(1) Models}},  {\em
  Phys. Rev.} {\bf D40} (1989) 1569--1585.

\bibitem{Czakon:1999ga}
M.~Czakon, J.~Gluza and M.~Zralek, {\it {Low-energy physics and left-right
  symmetry: Bounds on the model parameters}},  {\em Phys. Lett.} {\bf B458}
  (1999) 355--360 [\href{http://arXiv.org/abs/hep-ph/9904216}{{\tt
  arXiv:hep-ph/9904216}}].

\bibitem{Aad:2015owa}
{\bf ATLAS}, G.~Aad {\em et.~al.}, {\it {Search for high-mass diboson
  resonances with boson-tagged jets in proton-proton collisions at $\sqrt{s} =
  8$ TeV with the ATLAS detector}},
  \href{http://arXiv.org/abs/1506.00962}{{\tt arXiv:1506.00962}}.

\bibitem{Khachatryan:2014hpa}
{\bf CMS}, V.~Khachatryan {\em et.~al.}, {\it {Search for massive resonances in
  dijet systems containing jets tagged as W or Z boson decays in pp collisions
  at $ \sqrt{s} $ = 8 TeV}},  {\em JHEP} {\bf 08} (2014) 173
  [\href{http://arXiv.org/abs/1405.1994}{{\tt arXiv:1405.1994}}].

\bibitem{Khachatryan:2014gha}
{\bf CMS}, V.~Khachatryan {\em et.~al.}, {\it {Search for massive resonances
  decaying into pairs of boosted bosons in semi-leptonic final states at
  $\sqrt{s} =$ 8 TeV}},  {\em JHEP} {\bf 08} (2014) 174
  [\href{http://arXiv.org/abs/1405.3447}{{\tt arXiv:1405.3447}}].

\bibitem{Dobrescu:2015qna}
B.~A. Dobrescu and Z.~Liu, {\it {A W' boson near 2 TeV: predictions for Run 2
  of the LHC}},  \href{http://arXiv.org/abs/1506.06736}{{\tt
  arXiv:1506.06736}}.

\bibitem{Cheung:2015nha}
K.~Cheung, W.-Y. Keung, P.-Y. Tseng and T.-C. Yuan, {\it {Interpretations of
  the ATLAS Diboson Anomaly}},  \href{http://arXiv.org/abs/1506.06064}{{\tt
  arXiv:1506.06064}}.

\bibitem{Brehmer:2015cia}
J.~Brehmer, J.~Hewett, J.~Kopp, T.~Rizzo and J.~Tattersall, {\it {Symmetry
  Restored in Dibosons at the LHC?}},
  \href{http://arXiv.org/abs/1507.00013}{{\tt arXiv:1507.00013}}.

\bibitem{Gao:2015irw}
Y.~Gao, T.~Ghosh, K.~Sinha and J.-H. Yu, {\it {SU(2)xSU(2)xU(1) interpretations
  of the diboson and Wh excesses}},  {\em Phys. Rev.} {\bf D92} (2015), no.~5
  055030 [\href{http://arXiv.org/abs/1506.07511}{{\tt arXiv:1506.07511}}].

\bibitem{Dobrescu:2015yba}
B.~A. Dobrescu and Z.~Liu, {\it {Heavy Higgs bosons and the 2 TeV $W'$ boson}},
   {\em JHEP} {\bf 10} (2015) 118 [\href{http://arXiv.org/abs/1507.01923}{{\tt
  arXiv:1507.01923}}].

\bibitem{Coloma:2015una}
P.~Coloma, B.~A. Dobrescu and J.~Lopez-Pavon, {\it {Right-Handed Neutrinos and
  the 2 TeV $W'$ Boson}},  \href{http://arXiv.org/abs/1508.04129}{{\tt
  arXiv:1508.04129}}.

\bibitem{Deppisch:2015cua}
F.~F. Deppisch, L.~Graf, S.~Kulkarni, S.~Patra, W.~Rodejohann, N.~Sahu and
  U.~Sarkar, {\it {Reconciling the 2 TeV Excesses at the LHC in a Linear Seesaw
  Left-Right Model}},  \href{http://arXiv.org/abs/1508.05940}{{\tt
  arXiv:1508.05940}}.

\bibitem{Dev:2015pga}
P.~S. Bhupal~Dev and R.~N. Mohapatra, {\it {Unified explanation of the $eejj$,
  diboson and dijet resonances at the LHC}},  {\em Phys. Rev. Lett.} {\bf 115}
  (2015), no.~18 181803 [\href{http://arXiv.org/abs/1508.02277}{{\tt
  arXiv:1508.02277}}].

\bibitem{Bandyopadhyay:2015fka}
T.~Bandyopadhyay, B.~Brahmachari and A.~Raychaudhuri, {\it {Implications of the
  CMS search for $W_R$ on Grand Unification}},
  \href{http://arXiv.org/abs/1509.03232}{{\tt arXiv:1509.03232}}.

\bibitem{Hisano:2011cs}
J.~Hisano, K.~Ishiwata, N.~Nagata and T.~Takesako, {\it {Direct Detection of
  Electroweak-Interacting Dark Matter}},  {\em JHEP} {\bf 1107} (2011) 005
  [\href{http://arXiv.org/abs/1104.0228}{{\tt arXiv:1104.0228}}].

\bibitem{Hill:2011be}
R.~J. Hill and M.~P. Solon, {\it {Universal behavior in the scattering of
  heavy, weakly interacting dark matter on nuclear targets}},  {\em Phys.Lett.}
  {\bf B707} (2012) 539--545 [\href{http://arXiv.org/abs/1111.0016}{{\tt
  arXiv:1111.0016}}].

\bibitem{Hill:2013hoa}
R.~J. Hill and M.~P. Solon, {\it {WIMP-nucleon scattering with heavy WIMP
  effective theory}},  {\em Phys.Rev.Lett.} {\bf 112} (2014) 211602
  [\href{http://arXiv.org/abs/1309.4092}{{\tt arXiv:1309.4092}}].

\bibitem{Hill:2014yka}
R.~J. Hill and M.~P. Solon, {\it {Standard Model anatomy of WIMP dark matter
  direct detection I: weak-scale matching}},  {\em Phys.Rev.} {\bf D91} (2015)
  043504 [\href{http://arXiv.org/abs/1401.3339}{{\tt arXiv:1401.3339}}].

\bibitem{Aprile:2012nq}
{\bf XENON100 Collaboration}, E.~Aprile {\em et.~al.}, {\it {Dark Matter
  Results from 225 Live Days of XENON100 Data}},  {\em Phys.Rev.Lett.} {\bf
  109} (2012) 181301 [\href{http://arXiv.org/abs/1207.5988}{{\tt
  arXiv:1207.5988}}].

\bibitem{Akerib:2013tjd}
{\bf LUX Collaboration}, D.~Akerib {\em et.~al.}, {\it {First results from the
  LUX dark matter experiment at the Sanford Underground Research Facility}},
  {\em Phys.Rev.Lett.} {\bf 112} (2014), no.~9 091303
  [\href{http://arXiv.org/abs/1310.8214}{{\tt arXiv:1310.8214}}].

\bibitem{Cushman:2013zza}
P.~Cushman, C.~Galbiati, D.~McKinsey, H.~Robertson, T.~Tait {\em et.~al.}, {\it
  {Working Group Report: WIMP Dark Matter Direct Detection}},
  \href{http://arXiv.org/abs/1310.8327}{{\tt arXiv:1310.8327}}.

\bibitem{Mohapatra:1986bd}
R.~Mohapatra and J.~Valle, {\it {Neutrino Mass and Baryon Number
  Nonconservation in Superstring Models}},  {\em Phys.Rev.} {\bf D34} (1986)
  1642.

\bibitem{Nath:2001uw}
P.~Nath and R.~M. Syed, {\it {Analysis of couplings with large tensor
  representations in SO(2N) and proton decay}},  {\em Phys. Lett.} {\bf B506}
  (2001) 68--76 [\href{http://arXiv.org/abs/hep-ph/0103165}{{\tt
  arXiv:hep-ph/0103165}}]. [Erratum: Phys. Lett.B508,216(2001)].

\bibitem{Senjanovic:1978ev}
G.~Senjanovic, {\it {Spontaneous Breakdown of Parity in a Class of Gauge
  Theories}},  {\em Nucl. Phys.} {\bf B153} (1979) 334--364.

\bibitem{Senjanovic:1975rk}
G.~Senjanovic and R.~N. Mohapatra, {\it {Exact Left-Right Symmetry and
  Spontaneous Violation of Parity}},  {\em Phys. Rev.} {\bf D12} (1975) 1502.

\bibitem{Pati:1974yy}
J.~C. Pati and A.~Salam, {\it {Lepton Number as the Fourth Color}},  {\em Phys.
  Rev.} {\bf D10} (1974) 275--289. [Erratum: Phys. Rev.D11,703(1975)].

\bibitem{Mohapatra:1974gc}
R.~N. Mohapatra and J.~C. Pati, {\it {A Natural Left-Right Symmetry}},  {\em
  Phys. Rev.} {\bf D11} (1975) 2558.

\bibitem{Mohapatra:1980yp}
R.~N. Mohapatra and G.~Senjanovic, {\it {Neutrino Masses and Mixings in Gauge
  Models with Spontaneous Parity Violation}},  {\em Phys.Rev.} {\bf D23} (1981)
  165.

\bibitem{Belanger:2014vza}
G.~Bélanger, F.~Boudjema, A.~Pukhov and A.~Semenov, {\it {micrOMEGAs4.1: two
  dark matter candidates}},  {\em Comput. Phys. Commun.} {\bf 192} (2015)
  322--329 [\href{http://arXiv.org/abs/1407.6129}{{\tt arXiv:1407.6129}}].

\bibitem{Alloul:2013bka}
A.~Alloul, N.~D. Christensen, C.~Degrande, C.~Duhr and B.~Fuks, {\it {FeynRules
  2.0 - A complete toolbox for tree-level phenomenology}},  {\em Comput. Phys.
  Commun.} {\bf 185} (2014) 2250--2300
  [\href{http://arXiv.org/abs/1310.1921}{{\tt arXiv:1310.1921}}].

\bibitem{Roitgrund:2014zka}
A.~Roitgrund, G.~Eilam and S.~Bar-Shalom, {\it {Implementation of the
  left-right symmetric model in FeynRules/CalcHep}},
  \href{http://arXiv.org/abs/1401.3345}{{\tt arXiv:1401.3345}}.

\bibitem{Cirelli:2007xd}
M.~Cirelli, A.~Strumia and M.~Tamburini, {\it {Cosmology and Astrophysics of
  Minimal Dark Matter}},  {\em Nucl. Phys.} {\bf B787} (2007) 152--175
  [\href{http://arXiv.org/abs/0706.4071}{{\tt arXiv:0706.4071}}].

\bibitem{Ackermann:2015zua}
{\bf Fermi-LAT}, M.~Ackermann {\em et.~al.}, {\it {Searching for Dark Matter
  Annihilation from Milky Way Dwarf Spheroidal Galaxies with Six Years of Fermi
  Large Area Telescope Data}},  {\em Phys. Rev. Lett.} {\bf 115} (2015), no.~23
  231301 [\href{http://arXiv.org/abs/1503.02641}{{\tt arXiv:1503.02641}}].

\bibitem{Barducci:2015ffa}
D.~Barducci, A.~Belyaev, A.~K.~M. Bharucha, W.~Porod and V.~Sanz, {\it
  {Uncovering Natural Supersymmetry via the interplay between the LHC and
  Direct Dark Matter Detection}},  {\em JHEP} {\bf 07} (2015) 066
  [\href{http://arXiv.org/abs/1504.02472}{{\tt arXiv:1504.02472}}].

\bibitem{Boucenna:2015haa}
S.~M. Boucenna, M.~B. Krauss and E.~Nardi, {\it {Minimal Asymmetric Dark
  Matter}},  {\em Phys. Lett.} {\bf B748} (2015) 191--198
  [\href{http://arXiv.org/abs/1503.01119}{{\tt arXiv:1503.01119}}].

\bibitem{Fukugita:1986hr}
M.~Fukugita and T.~Yanagida, {\it {Baryogenesis Without Grand Unification}},
  {\em Phys. Lett.} {\bf B174} (1986) 45.

\bibitem{Davidson:2008bu}
S.~Davidson, E.~Nardi and Y.~Nir, {\it {Leptogenesis}},  {\em Phys. Rept.} {\bf
  466} (2008) 105--177 [\href{http://arXiv.org/abs/0802.2962}{{\tt
  arXiv:0802.2962}}].

\bibitem{Davidson:2002qv}
S.~Davidson and A.~Ibarra, {\it {A Lower bound on the right-handed neutrino
  mass from leptogenesis}},  {\em Phys. Lett.} {\bf B535} (2002) 25--32
  [\href{http://arXiv.org/abs/hep-ph/0202239}{{\tt arXiv:hep-ph/0202239}}].

\bibitem{Arbelaez:2013nga}
C.~Arbeláez, M.~Hirsch, M.~Malinský and J.~C. Romão, {\it {LHC-scale
  left-right symmetry and unification}},  {\em Phys. Rev.} {\bf D89} (2014),
  no.~3 035002 [\href{http://arXiv.org/abs/1311.3228}{{\tt arXiv:1311.3228}}].

\end{thebibliography}

 \providecommand{\href}[2]{#2}\begingroup\raggedright\endgroup

\end{document}